 \documentclass[final,3p,times,authoryear]{elsarticle}
\usepackage{amssymb}
 \usepackage{amsthm}
 \usepackage{amsmath}
 \usepackage{color}
 \usepackage{amsmath}
\usepackage{siunitx}
\usepackage{todonotes}
\usepackage{framed} % Framing content
\usepackage[]{algorithm2e}

\renewcommand{\cite}[1]{~\citep{#1}}

\journal{arXiv}

\begin{document}

\begin{frontmatter}



\title{The Nature of Human Settlement: Building an understanding of high performance city design.}


\author[melb]{Kerry~A.~Nice\corref{cor1}\fnref{label1}}
\ead{kerry.nice@unimelb.edu.au}
\author[melb]{Gideon D.P.A. Aschwanden\fnref{label1}}
\author[melb]{Jasper S. Wijnands}
\author[melb,sunshine]{Jason Thompson}
\author[melb]{Haifeng Zhao}
\author[melb,eng]{Mark Stevenson}

\cortext[cor1]{Principal corresponding author}
\address[melb]{Transport, Health, and Urban Design Hub, Faculty of Architecture, Building, and Planning, University of Melbourne, Victoria 3010, Australia}
\address[eng]{Melbourne School of Engineering; and Melbourne School of Population and Global Health, University of Melbourne, Victoria, Australia.}
\address[sunshine]{Centre for Human Factors and Sociotechnical Systems, University of the Sunshine Coast, Australia.}
\fntext[label1]{These authors contributed equally to this work}

\begin{abstract}

In an impending urban age where the majority of the world's population will live in cities, it is critical that we improve our understanding of the strengths and limitations of existing city designs to ensure they are safe, clean, can deliver health co-benefits and importantly, are sustainable into the future. To enable this, a systematic and efficient means of performing inter- and intra-city comparisons based on urban form is required. Until now, methods for comparing cities have been limited by scalability, often reliant upon non-standardised local input data that can be costly and difficult to obtain. To address this, we have developed a unique approach to determine the mix, distribution, and composition of neighbourhood types in cities based on dimensions of block size and regularity, sorted by a self-organising map. We illustrate the utility of the method to provide an understanding of the underlying city morphology by overlaying spatially standardised city metrics such as air pollution and transport activity across a set of 1667 global cities with populations exceeding 300,000. The unique approach reports associations between specific mixes of neighbourhood typologies and quantities of moving vehicles (r=0.97), impervious surfaces (r=0.86), and air pollution levels (aerosol optical depth r=0.58 and NO$_{2}$ r=0.57). What this illustrates, is that this unique approach can identify the characteristics and neighbourhood mixes of well-performing urban areas while also producing unique `city fingerprints' that can be used to provide new metrics, insights, and drive improvements in city design for the future.

\end{abstract}


\end{frontmatter}

\section{Introduction}
The world's urban areas are rapidly growing in size, density, and in importance with more than half the world's population living in cities; this is projected to increase to two-thirds by 2050\cite{UNDESA2019}. Beyond population growth, internal migration from rural areas and the concentration of international arrivals into cities is further driving this urbanisation process\cite{Raymer2018}. These accelerating changes and the growing need to accommodate increasing urban populations presents some key challenges but also opportunities. Poorly managed urbanisation can impact urban residents through increased pollution\cite{Stevenson2016,Sallis2016,Landrigan2017}, urban heat\cite{Coutts2012,Bowler2010}, urban sprawl\cite{Frank2000,Bettencourt2010}, and social isolation\cite{Vlahov2002}. However, urbanisation can also promote economic activities and innovation, increased employment opportunities, and community building through centralised locations, with economies of scale reducing the need for sprawling infrastructure\cite{Kuhnert2006,Bettencourt2007,Lobo2013}.

Managing this process requires data and knowledge about how cities work. However, we lack an objective method to compare cities around the world and especially lack a systematic quantitative method to identify different neighbourhoods and perform inner-city comparisons\cite{Louf2014a}. This leads to repeated reinvention of solutions for problems that occur globally and limits the possibility to learn from previous examples. Understanding and comparing cities has been done many times where crude aggregates of population density or GDP measures are juxtaposed. Our framework allows us to compare a single element, as small as neighbourhoods of cities, across the globe.

To discover how different cities work, previous research has been conducted through a number of methods to categorise cities and analyse underlying processes. Analysing the basic structural elements of road networks and urban blocks, often the most long lasting part of an urban area, can provide clues as to the processes under which city development occurs\cite{Porta2006a,Strano2012}. In addition, the road structure can point to the dominant modes of transportation and governance systems underlying each urban area, grid structures reflecting a top-down planning system\cite{Crouch1977,Courtat2011} while T-shaped crossings point to more disorderly\cite{Jacobs1961} self-organised organic growth\cite{Cardillo2006}. Division of large land blocks (often originally agricultural land) can follow an evolutionary progression, either to medium sized manufacturing or smaller residential plots\cite{Fialkowski2008}. In addition, studies show that areas unconstrained by adjoining villages or topography are generally and most efficiently subdivided into smaller grids (i.e. regular rectangular plots)\cite{Strano2012}. While this research based on urban blocks and streets reveal insights into cities and how the resulting built environment influences a wide range of outcomes, including health, transport, and economic opportunity, they are limited by scalability and observer dependence. In addition, validation and applicability of these methods to all types of world-wide cities are limited by the ability to collect urban information that applies to all cities.
 
To overcome these difficulties, and building on recent advances in computing power, artificial intelligence, and the wide availability of urban imagery, new approaches have been created to discover unique characteristics of cities and how cities function. Large numbers of geo-tagged photos have been used to detect patterns of urban usage and public perception of a number of areas' functional and social attributes\cite{Liu2016,Zhou2014a}. Place Pulse, a database of urban imagery using crowd-sourced classifications (including safety, beauty, and liveliness) has attempted to quantify perceptions of urban areas\cite{Dubey2016,Naik2014} and inequality\cite{Salesses2013}. Doersch\cite{Doersch2012} used a large number of geo-localised street level images to discover common visual features across a number of cities. Enabled by remote sensing data, night-time light data has been used to categorise cities into stages of urbanisation and levels of economic activities\cite{Zhang2013}. 

Urban characteristics (road geometry, building dimensions and heights, and vegetation heights) have also been used to classify cities into typologies of differing periods of historical design and urban planning (i.e. 19th Century, 1950s, 1970s, etc.)\cite{Hermosilla2014}. The connection between the physical and topological structure of the road network in cities to the structural sociology field, transportation and economics, has been drawn by the `space syntax' community and Hillier\cite{Hillier1996} that established a correlation between configurations of urban forms and variations of human interactions within it. 

Most methods described above require some amount of subjective classification of local input data; the quality and availability of which can vary widely across collection or political districts. Existing empirical methods highlight these mechanisms by evaluating the street network typology\cite{Hillier1989}, but neglect their geometrical expanse and their function as places to stay. A method cannot rely solely on topology but needs to incorporate the urban geometry\cite{Louf2014a}. 

Our proposed method uses block size and regularity information within a neighbourhood to comprehensively explore the characteristics of cities across selected domains. The fundamental nature of city blocks, defined as the area bounded by surrounding streets, can be read as a simplified schematic view of the city\cite{Southworth2013}, highlighting both the structure and organisation of the city, as well as the process of the urban development and morphology. This makes the city block the most used and accessible urban element for urban analysis and basic common elements in city design typologies and theory. Blending theory with globally available datasets at scale can enable new insights into the form and function of the world's cities. This method allows us to find the mix of commonalities and unique elements, a `fingerprint' of each city, to understand what about each city works and doesn't work, an understanding essential to best manage the world's rapid urbanisation for health and well-being.

\section{Methods}\label{sec:Methods}

%\section{Methods}\label{sec:Methods}

\subsection{Map imagery sampling.}\label{sec:methods2}
The concept employed in this study was to sample maps of individual city sections, calculate block size and regularity of each section, and then use a self organising map (SOM) to organise the images into different urban types. All cities with populations greater than 300,000 people\cite{UN2014} were selected for analysis. Map imagery from Google Maps\cite{GoogleStatic2017} was used to provide globally consistent data. 

A two-stage sampling approach was applied to each city. As no standardised urban boundaries are available for all the cities evaluated in this study, a methodology had to be developed to define these. Firstly, a sampling area extending 1.5 km from the identified city centroid\cite{UN2014} was set as a baseline. Then the sampling radius $r$ (km) was scaled, increasing by a power of 0.85 to the proportional increase in population size based on Barthelemy\cite{Barthelemy2016} in 

\begin{equation}
r = \sqrt{ \frac{28.27}{\pi} \left( \frac{p}{300,000}  \right) ^{0.85} }
\end{equation}

Standardising the sampling area in this manner avoided socio-political discrepancies relating to a city's `true' (political) boundary and captured differences in population density and shape between small (e.g., Wellington, New Zealand; Izmit, Turkey) and global mega-cities (e.g., Tokyo, Japan;  Delhi, India). Location sampling areas were adjusted for the earth's curvature\cite{Sinnott1984}. Large water-bodies (e.g., oceans but not coastlines) were removed from the sampling area, as they were not indicative of urban form. 

\begin{figure}
    \centering    
\includegraphics[trim={0 0 0 0},clip,scale=0.15]{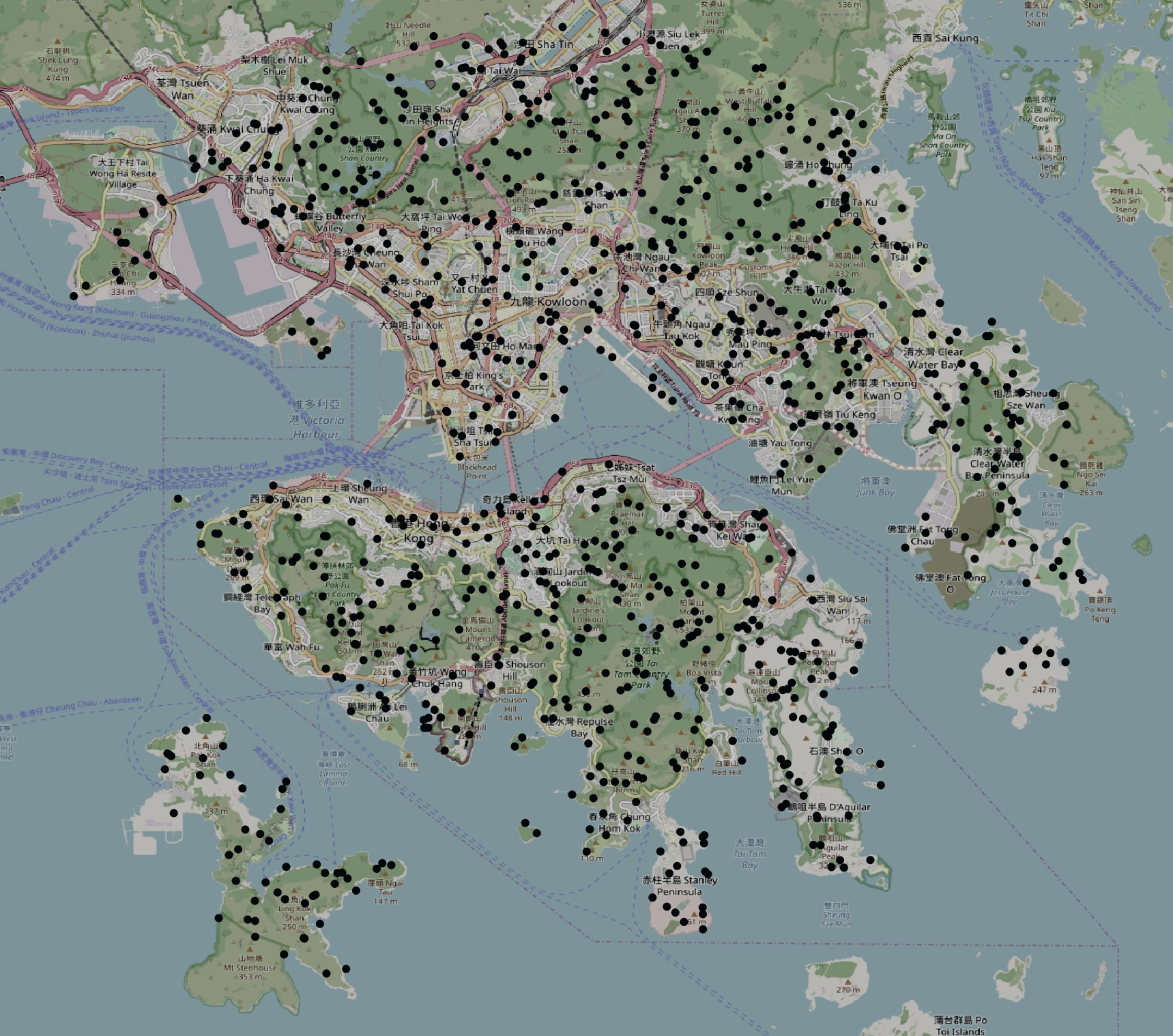}  
\caption{\bf Sampling locations for map imagery (from Hong Kong).}    
 \label{fig:parissample}  
\end{figure}

These procedures result in a population and water body-adjusted circular area centred on the city's central coordinates, intended to capture the widest extent of each city while minimising the amount of non-urban locations. For example, Figure \ref{fig:parissample} shows the resulting sampling locations used in collecting imagery for Hong Kong. 

\subsection{Map imagery source.}\label{methodsimagery}

\begin{figure}
    \centering    
% \includegraphics[scale=0.8]{Images/SampleTraining.png}   
%\frame{
%\includegraphics[page=5,trim={80 365 82 365},clip,scale=0.65]{BlockTypologies_Figures2.pdf} 
\includegraphics[trim={0 0 0 0},clip,scale=0.15]{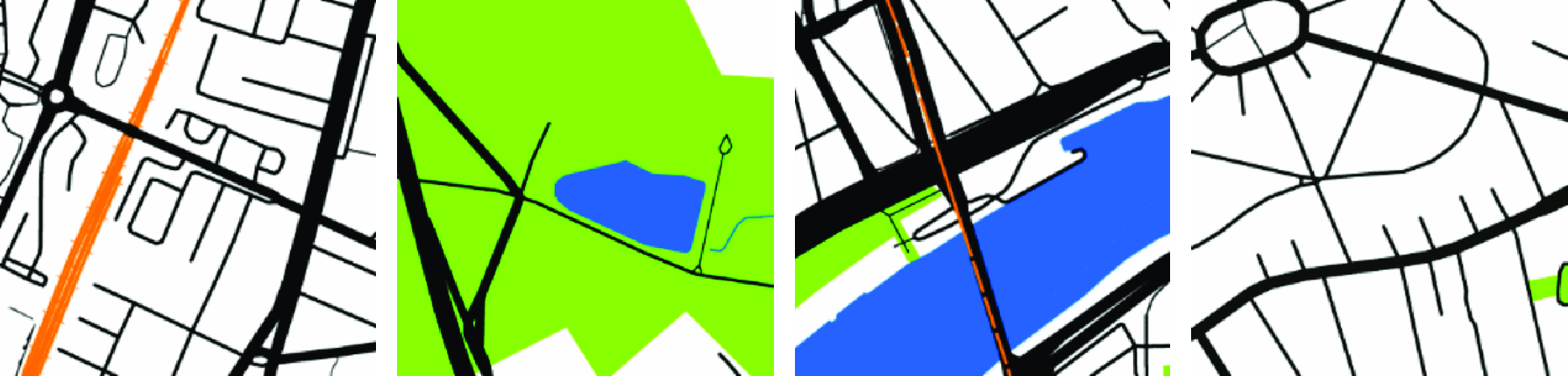}  
%}
\caption{\bf Four sample Google Maps training data images (from Paris, France)\cite{GoogleStatic2017}.}    
 \label{fig:maps}  
\end{figure}

320$\times$320 pixel sized map images were sampled for 1692 cities using a zoom level of 16 (covering 750$\times$750m at the equator and down to 335$\times$335m at higher latitudes) using a custom style defined with the Google Static Maps API\cite{GoogleStatic2017} (see Figure~\ref{fig:maps} for examples of Paris, France). To ensure each map covers the same area, each image was cropped and resized before processing. The sampled city at the highest latitude was at 64 degrees north, so each image was cropped and resized to include a region of 335$\times$335m. 

The maps provide a high-level abstraction of road (black) and public transport (orange) networks, green space (green), and water bodies (blue). Any remaining space is coded white. Inconsistent map imagery from 25 South Korean cities (due to South Korean government restrictions on map data\cite{Badalge2018}) was removed from the dataset, reducing the number of cities to 1667. 1000 maps were sampled per city. The total dataset consists of nearly 1.7 million images.

\subsection{Calculating block size, regularity, and colour counts.}\label{methodscalc}

Block size, regularity, and colour counts were calculated for each sampled image with Algorithm \ref{alg:floodfill}, using the Java 8 AWT toolkit\cite{Oracle2018}:

\begin{algorithm}[H]\label{alg:floodfill}
\SetAlgoLined
\KwResult{Histogram of region sizes and region regularity for a single image}
 Using latitude of image sample, crop the image to represent 335$\times$335 meters;\\
 Resize image back to 320$\times$320 pixels;\\
 Start at top left point of image;\\
 \While{White pixels are found}{
  Floodfill area using boundaries of all non-white colours (i.e. black, green, blue, orange);\\
  Count pixel size of region;\\  
  Construct a the smallest bounding box of the cloud of points in the region using the Fast Convex Hull algorithm\cite{Javagl2017,GoogleArchive2011};\\
  Use the difference of counted pixels between the bounding box size and the region size as measure of irregularity;\\
  Add size and regularity counts to corresponding (pre-specified) histogram bin;\\
  Locate next white pixel by iterating across rows and columns;\\
 }
 Count percentage of blue, orange, green, black, and white pixels in each image.\\
 Combine the two size and regularity histograms along with colour counts into a single histogram vector to be used in the SOM.\\
 \caption{Calculation of histograms of block sizes and regularity}
\end{algorithm}

\begin{figure}
\centering    
 \includegraphics[trim={0 0 0 0},clip,scale=0.25]{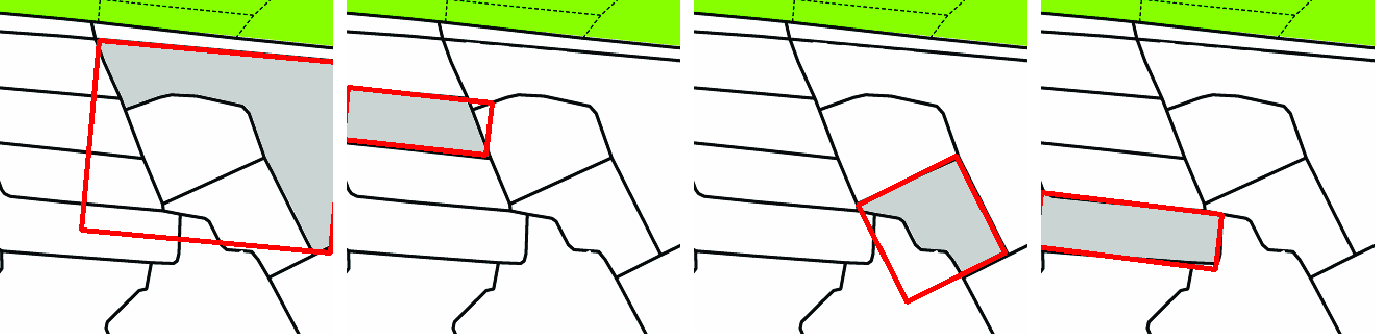}  
\caption{\bf Results of flood filled city blocks showing flood fills of each individual region to determine region size (count of pixels in grey). Differences between region size and pixel counts within bounding boxes (outlined in red) are used as a measure of regularity.}    
 \label{fig:floodfilled}  
\end{figure}

Samples of size floodfills and regularity floodfills are shown in Figure \ref{fig:floodfilled}. Sample histograms used in the SOM are shown in Figure \ref{fig:mapsandHist}.

\subsection{City size and regularity histograms.}\label{methodshist}

Using the calculated counts, two vectors were constructed for each image, one each for block size and block regularity. The vectors were sorted into 15 histogram bins (the number of bins determined by Sturges' formula\cite{Sturges1926}, $\lceil \log_{2}n \rceil +1$, with $n$ being the number of data points). To reduce the clumping of data in the first bin, bins of increasing sizes were used to spread this data across all bins. The first bin starts with a size boundary of 1 and each following bin has a boundary of the current bin boundary times a multiplier. A multiplier of 2.3 was used to fit the maximum count size (320$\times$320 pixels = 102400) into the 15 bins.

\begin{figure}
\centering    
%\frame{
 %\includegraphics[page=7,trim={62 395 63 384},clip,scale=0.95]{BlockTypologies_Figures2.pdf} 
 \includegraphics[trim={0 0 0 0},clip,scale=0.20]{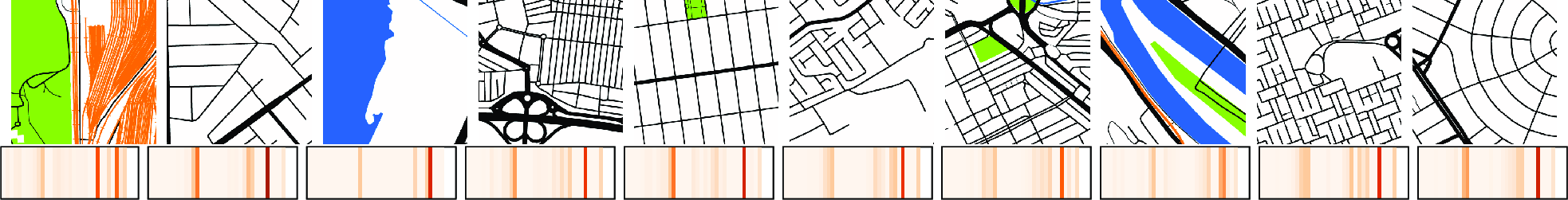}  
%   }
\caption{\bf Samples of map regions (top) and resulting histograms (bottom). Region size, regularity, and colour counts are joined into a combined histogram vector, with size frequencies in the first 15 bins, regularity in the second 15 bins and colour pixel counts in the remaining 5 bins.}    
 \label{fig:mapsandHist}  
\end{figure}

The resulting histograms for sample map regions are shown in Figure \ref{fig:mapsandHist}. Histograms input into the SOM were constructed by combining the 15 bins of region size frequencies (on the left side) with the 15 bins of region regularity frequencies (second 15 bins) into a single histogram vector. In addition, the 5 colour pixel percentages are appended to the end of the histogram vector. Finally, the vector values are normalised into a range of 0 to 1.

\subsection{Sorting map histograms in the self organising map.}\label{methodscluster}
The SOM methodology\cite{Kohonen1982} is a data driven technique that transforms a multi-dimensional data source into a lower dimensional space, commonly a two-dimensional map, while keeping the relative proximity of two datapoints intact. The distance in the lower dimensional representation is therefore a similarity index, calculated as the euclidean distance, of the higher dimensional space. Each point in the two-dimensional map has location (x,y) and is associated with a vector of values from the multi-dimensional space.

SOM is a generic, objective and robust methodology that has been deployed in many domains and is used for the visualisation of multi-dimensional data and data exploration\cite{Koleheimen2004}. This methodology was chosen for its ability to create two-dimensional maps of smoothly changing patterns from the original high-dimensional space. Additionally, the SOM map spans the extremes observed in the original data and allows for investigation on how the data is distributed, potential paths between two observations. 

The 1.7 million map histograms with 35 dimensions were the initial data space used to train the two-dimensional SOM. After the randomised initialisation of the 100x100 nodes of the SOM, a random selection of 5.4 million data points from the initial data space were used to transform the two-dimensional SOM to match it. This iterative process locates nodes that are similar to the training vector and morphs the values of the SOM nodes towards the training values. The number of iterations were determined by the minimum number of iterations to reach the greatest continuity among clusters. Training was stopped when the clusters started becoming discontinuous, checking every 150,000 iterations for a rapid increase in number of clusters. However, experimentally, we found that once a sufficient number of iterations were run, subsequent iterations only had small impacts on the results. 

This training is subjugated to a decay function for both magnitude (learning decay, $L_{i}$) and distance (radius decay, $D_{R}$) in the SOM. Radius decay was calculated using

\begin{equation} 
D_{R} = r_{0} e^{-\frac{i}{n} \log _{10} (r_{0})}
\end{equation}
where radius ($r_{0}$) is 50 (the width and height of the SOM was 100) and the current training iteration $i$ (of total iterations $n$ of 3.2 million). Learning decay is calculated as
\begin{equation} 
L_{i} = L_{0}  e^{-\frac{i}{n}}
\end{equation}
where learn rate ($L_{0}$) = 0.05.

After the SOM was trained, each map histogram was classified to find the closest matching node in the SOM. The underlying imagery of the resulting trained SOM was visualised by tiling representative map images from each node (x,y) point in the SOM in Figure \ref{fig:somresults}. Areas with black have no associated map segments with that particular node's (x,y) location (about 5000 nodes). Most nodes are associated with multiple map segments that have similar characteristics. One notable outlier exists that accumulates more than 60,000 map segments of (0,99) with 385,967 maps that are all or nearly all white. Approximately 20 nodes contain 60,000 to 10,000 maps, about 200 contain 10,000 to 1000 maps, 5700 nodes contain 1000 to 1 map.

%0	99	385967
%99	8	58542
%46	99	50585

The node's (x,y) locations were encoded into RGB colour codes using a Java 8 port of Color2D\cite{Jackle2017,Steiger2015}. These colours were used in plotting the (x,y) typologies in QGIS\cite{QGIS2009}.

\subsection{KernelDensityEstimator2D.}\label{kerneldensity}

To create the city fingerprints, kernel density plots were made of the SOM x,y locations for each city using the KernelDensityEstimator2D (a bi-variate kernel density smoother for data) module from the Bayesian Evolutionary Analysis by Sampling Trees (BEAST) software package \cite{Suchard2018} with a smoothed grid size of 100.

% KernelDensityEstimator2D kde = new KernelDensityEstimator2D(dataXa, dataYa, null, 100, null); // 100 is smoothed grid size

\subsection{Aerosol Optical Depth (AOD) Dataset}\label{aod}
Two near-identical MODIS (Moderate Resolution Imaging Spectroradiometer) instruments are mounted upon the sun-synchronous polar-orbiting NASA satellites Terra and Aqua; these missions have over-pass times of approximately 10:30am and 1:30pm local time and were launched in 1999 and 2004, respectively. The MODIS resolution is 10 km $\times$ 10 km at nadir. The MOD04 and MYD04 L2 retrievals (V006) were downloaded and gridded to 0.05$^\circ$ x 0.05$^\circ$ (grid-spacing of approximately 5.6km). A range of different retrievable fields are available of which we used "Optical\_Depth\_Land\_And\_Ocean" (see Table B1 of Levy et al. (2013)\cite{Levy2013}), which represents a compromise between quality and coverage. These data were averaged to a monthly temporal resolution on this grid, and the number of MODIS pixels contributing to each averaged value were recorded. At each grid-point, a time-series decomposition into seasonal, trend and irregular (STL) components was applied\cite{Cleveland1990}. A slight modification to the code of the STL algorithm in the R stats package\cite{RCoreTeam2015} was made so that data were weighted by the number of observations involved in creating these observations. The trend component is estimated as a smooth function (via the locally weighted scatterplot smoothing, or LOESS, algorithm of Cleveland et al. (1992)\cite{Cleveland1992}), however the trend window parameter (defining the smoothing length scale for the trend term) was set sufficiently high that this term was effectively a linear term. From this, we derived the mean concentration at each grid-point accounting for long-term trend and seasonal fluctuations.

\subsection{NO$_{2}$ Dataset}\label{no2}
The tropospheric-column NO$_{2}$ data were derived from the TEMIS (Tropospheric Emission Monitoring Internet Service) OMI (Ozone Monitoring Instrument) tropospheric-column NO$_{2}$ (tcNO$_{2}$) database\cite{Boersma2007a}. Monthly gridded averages at 0.125$^\circ$ $\times$ 0.125$^\circ$ resolution (a grid-spacing of roughly 13km) were downloaded from the TEMIS website. These are based on the Level-2 tropospheric-column retrievals. While this provides little vertical information, the tcNO$_{2}$ product shows a good correlation with surface NO$_{2}$ concentrations, when averaged over a sufficient period. The OMI is mounted aboard NASA's  sun-synchronous, polar-orbiting satellite Aura (launched July 2004). In normal operation, the OMI pixels are 13 km $\times$ 24 km at their finest (i.e. at nadir). The combination of the instruments wide swath (spanning about 2600 km) and the 14 orbits daily provides a global coverage every day, however retrievals are not possible everywhere, mainly due to clouds. Airborne aerosols and surface albedo are other significant sources of uncertainty. The monthly tcNO$_{2}$ data from 2005-2016 (complete years) were averaged at their native resolution.

\subsection{City fractions dataset}\label{gsvdata}
Middel et al. (2019,2018)\cite{Middel2019,Middel2018} derived fractional breakdowns from Google Street View (GSV) panorama images for 65 million locations across 75 cities of six urban form classes of sky, trees, buildings, impervious surfaces, pervious surfaces, and non-permanent objects (i.e. moving vehicles). Data for each location was indexed by city, latitude, and longitude. 

\subsection{City correlations}\label{correlations}
Mean averages for each city were generated using the same centroid area as in the "Map imagery sampling" section above, however locations were sampled at a 400$\times$400m resolution instead of the 1000 randomly selected locations. Using these locations, mean values of AOD and NO$_{2}$ were calculated for 1667 cities. A second set of city mean values were generated for 34 cities that matched the cities sampled for this study using the city fractions dataset.

Next, mean averages of pollution (AOD and NO$_{2}$) and of urban form fractions were calculated for each SOM(x,y) location. The latitude and longitude of each point (for the 1.7 million points) was used to look up an individual value from the gridded pollution datasets. For the city fractions dataset, latitude and longitude locations were matched to 0.0001 degree, for a total of approximately 8500 locations.

Finally, mean averages were calculated using the SOM(x,y) averages for the 1000 locations for 1667 and 34 cities for pollution and urban form respectively. Then the cor() function, using "pairwise.complete.obs" and "pearson", from the R stats package\cite{RCoreTeam2015} was used to find correlations between the two sets of computed city averages for the pollution and city fractions datasets.

\section{Results}

This project provides a means to examine all built environment designs from the largest cities around the world with greater than 300,000 population using an unsupervised sorting algorithm on maps segments of 300$\times$300m sampled from these cities. This provides insights at an unprecedented granularity, uncovering the scale of the neighbourhood, and the nature of cities. We have found unique individual neighbourhood types that can have twins in different cities across the globe. We also show the extent to which individual cities are heterogeneous within their bounds and the variation in distribution of neighbourhood types they display.

\begin{figure}
\centering    
\includegraphics[trim={0 0 0 0},clip,scale=0.17]{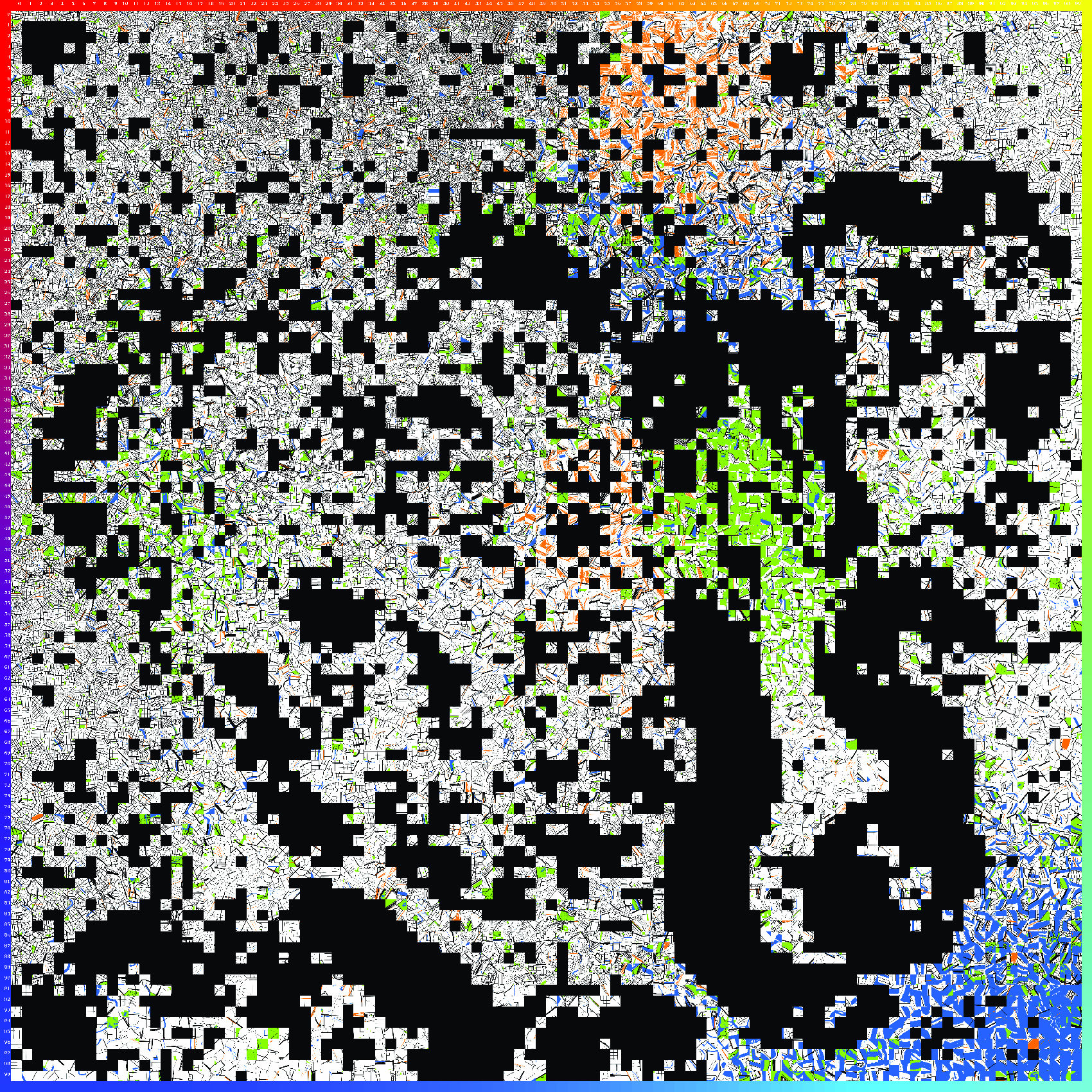}  
\caption{\bf  A visualisation of the 2-dimensional 100$\times$100 SOM trained with 1.7 million map images from 1667 cities.  Each (x,y) point shows a representative image associated with each node while nodes without associated images are shown in black. Border shows colour coding scheme for SOM (x,y) locations used in Figure \ref{fig:citylocations}.}    
 \label{fig:somresults}  
\end{figure}

The methodology we present uses data on land use and block size extracted from maps and deploys a Self Organising Map (SOM) algorithm to sort them into a tractable array. This method is both universally applicable and extendible to incorporate additional datasets available while scaling linearly. The output of the SOM (Figure \ref{fig:somresults}) contains an organised representation of all currently existing neighbourhood typologies in the urban areas where a large proportion (2.2 billion) of humans live in the world. The SOM sorts these neighbourhoods according to their typological elements, finding similarities in a high-dimensional space and projecting them into two-dimensions. Due to the stability of the physical structure of cities, undergoing long term small incremental changes\cite{Wegener1986}, this collection represents a collective representation of the built history of human settlements.

\begin{figure}
\centering  
\includegraphics[trim={0 0 0 0},clip,scale=0.13]{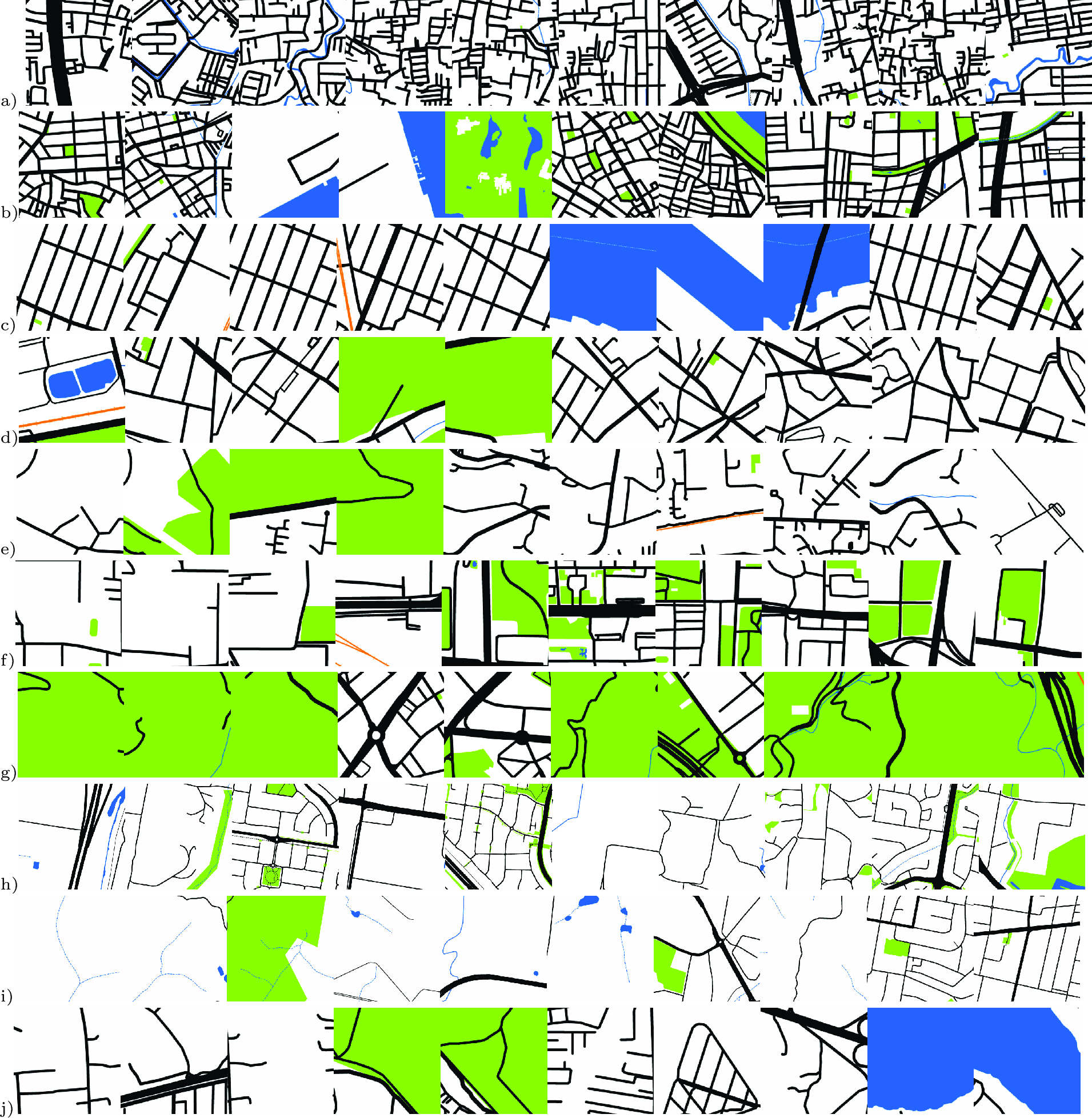}  
\caption{\bf Sample representative maps for top SOM (x,y) locations for cities 
a) Jakarta,
b) Tokyo, 
c) New York, 
d) Paris,
e) Nairobi,
f) Beijing, 
g) Barcelona, 
h) Melbourne, 
i) Sydney, and
j) Bras\'{i}lia.  
} \label{fig:TopSomMaps}  
\end{figure}

\begin{figure}
\centering   
\includegraphics[trim={0 0 0 0},clip,scale=0.17]{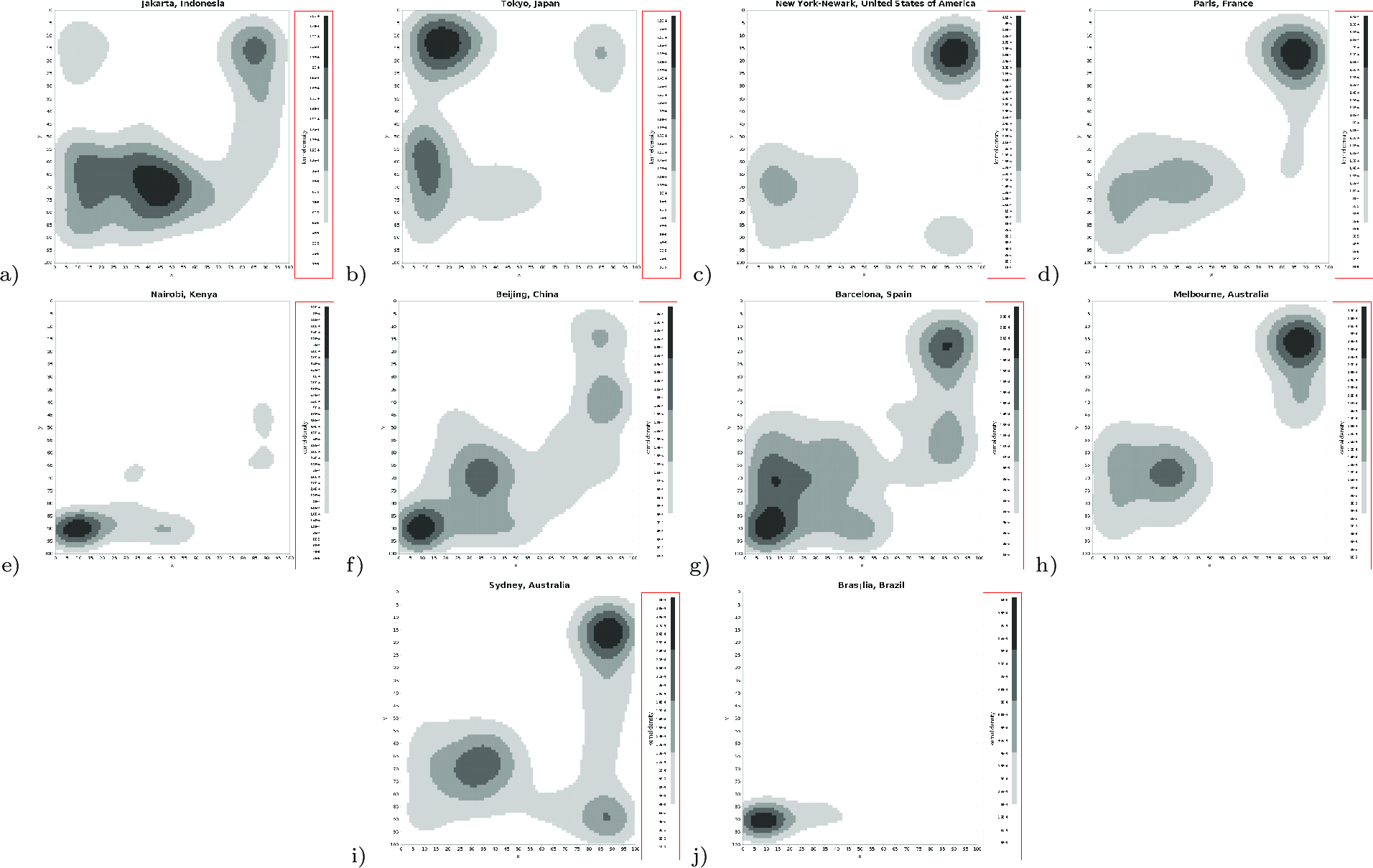}  
\caption{\bf City fingerprints generated by kernel density maps of SOM (x,y) locations for cities 
a) Jakarta,
b) Tokyo, 
c) New York, 
d) Paris,
e) Nairobi,
f) Beijing, 
g) Barcelona, 
h) Melbourne, 
i) Sydney, and
j) Bras\'{i}lia.  
}    
 \label{fig:kernel}  
\end{figure}

\begin{figure}
\centering    
%\frame{
%\includegraphics[page=1,trim={62 322 74 322},clip,scale=0.90]{BlockTypologies_Figures3.pdf} 
\includegraphics[page=1,trim={0 0 0 0},clip,scale=0.20]{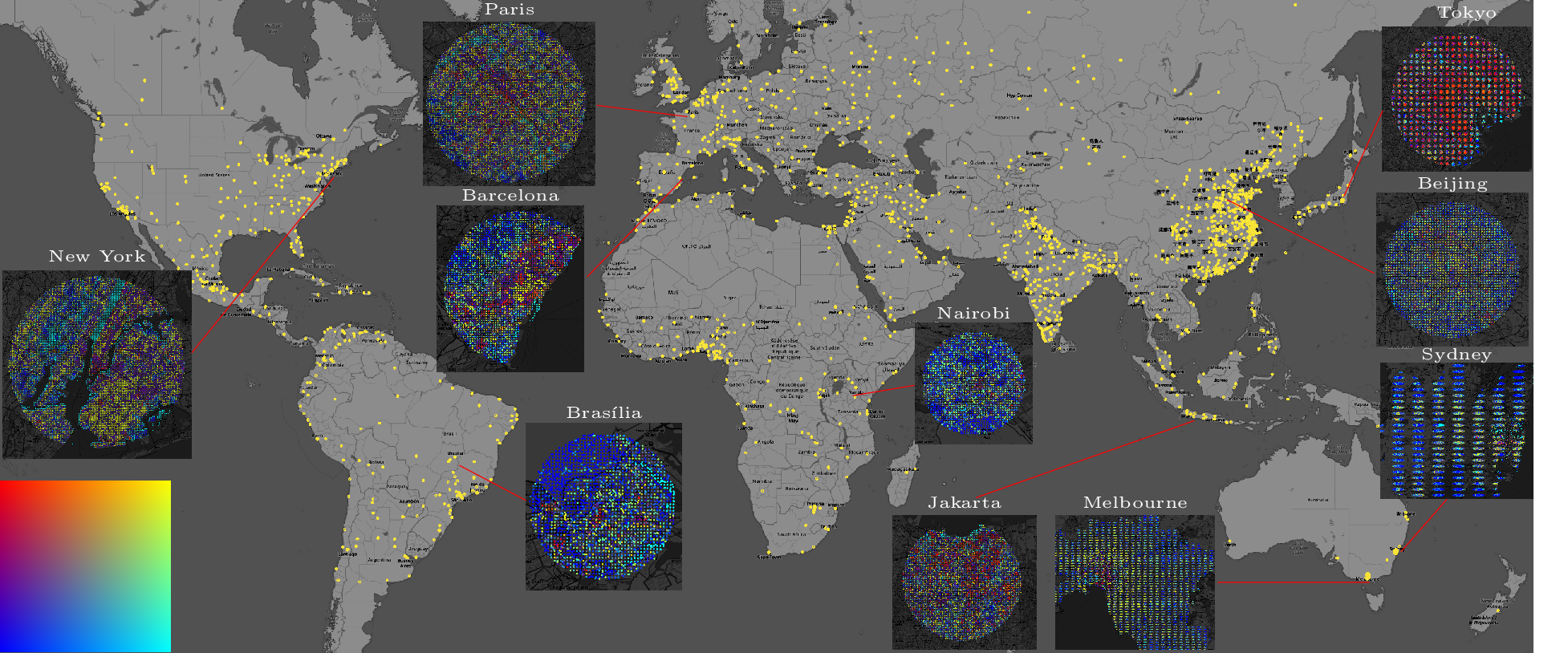}  
%   }
\caption{\bf Sampled world cities with inserts showing detail of New York, Paris, Barcelona, Bras\'{i}lia, Nairobi, Jakarta, Melbourne, Tokyo, Beijing, and Sydney. City detail maps use the same SOM (x,y) location colour scheme as the border of Figure \ref{fig:somresults} and of colour map insert image (lower left).}    
 \label{fig:citylocations}  
\end{figure}

A look at the range of neighbourhood typologies from individual cities shows that many cities are an eclectic mixture of different neighbourhood typologies (Figure \ref{fig:TopSomMaps}) but all are built using common elements. An individual city's uniqueness (in this case, the city's structure through its blocks and streets), its `fingerprint', is reflected in an individual mix of neighbourhood typologies (Figure \ref{fig:kernel}). These individual fingerprints can be used to compare cities around the world at a granularity of individual 300$\times$300m neighbourhoods. The uniqueness of cities are thus derived from a subtle mix and distribution of common elements (Figure \ref{fig:citylocations}).

As an illustration of the utility of this technique, we calculated correlations between mean average values of pollution and elements of urban form calculated for each city compared to averages calculated from a weighted average of the mix of SOM(x,y) locations for each city. The datasets used are a global gridded pollution datasets of Aerosol Optical Depth (AOD) and NO$_{2}$ and dataset of fractions of six urban form classes of sky, trees, buildings, impervious surfaces, pervious surfaces, and non-permanent objects (i.e. moving vehicles) at 65 million locations in 70 cities from Middel et al. (2019,2018)\cite{Middel2019,Middel2018}. These values are reported in Table \ref{table:correlations}.

\begin{table}
\caption{Correlations between mean average values by city and by (x,y) location within the SOM.}
\label{table:correlations}
\begin{tabular}{  | c | c |}
\hline  \textbf{Parameter} & \textbf{Correlation value}\\ \hline
Movable objects fraction& 0.97 \\ \hline
Impervious surfaces fraction& 0.86 \\ \hline
Sky fraction& 0.75 \\ \hline
Building fraction& 0.56 \\ \hline
Mean AOD& 0.58 \\ \hline
Mean NO$_{2}$&0.57 \\ \hline
\end{tabular}
\end{table}

%       row    column       cor     p
%aodAquaObs     xyAodAqua     0.5833058 0
%aodTerraObs     xyAodTerra     0.5678217 0
%no2Obs      xyNo2         0.5714648 0 
%        SVFByCity          SVFByXY  0.695137382 7.142315e-06
% imperviousByCity   imperviousByXY  0.858961761 8.040502e-119
%     movingByCity       movingByXY  0.973403215 0.000000e+00
%   perviousByCity     perviousByXY  0.432223290 1.068619e-02
%        skyByCity          skyByXY  0.754313707 2.576413e-07
%      treesByCity        treesByXY  0.330958635 5.588998e-02 
% buildingByXY          0.5558685  1.000000e+00

%  /home/kerryn/git/2019-02-BlockTypologyData/Code/output/somClassifiedStops_5400000_1800KImg10000KItr_20190606_no_green.csv 

\section{Broader Implications}

Our contribution is a method that allows global inter-comparisons of neighbourhoods while highlighting that cities are constructed using fundamental particles; neighbourhood typologies. The implications are both fundamental and practical. The results provide for the first time a comprehensive overview of the nature of human settlements, their distribution and prevalence. This allows improvements to other systems such as health, transportation, and employment that are built on these fundamental components. The system can provide guidance for designers, engineers, stakeholders and policy makers by harnessing insights from a comparison across the globe. 

As shown in the results, we find strong correlations between different mixes of neighbourhood typologies derived through block size and regularity and the morphology and composition (through fractions of movable vehicles, sky view, buildings and impervious surfaces) of these areas. In addition, we find correlations between the mix of neighbourhood typologies and the performance of the city in terms of pollutants (AOD and NO$_{2}$).

Our method has highlighted that cities are not unique and that individual neighbourhoods can be compared across continents. The findings also indicate that there is a common structure and that city typologies should be built across cities rather than within, that city centres are more comparable to other centres than to other elements of the same city. It also allows us now to compare cities on a global scale and investigate the nature of human settlements. In contrast to previous methods which were limited by the available data from a handful of cities, this method is capable of spanning the globe and can accommodate new datasets as they arise. For example, the link between urban form and vehicle emissions has been shown on an aggregate level\cite{Frank2000}. This can now be investigated at a neighbourhood scale. Similarly, transportation mode choices are associated with block size and accessibility that are primary enabler / inhibitor for residence using active modes of transportation. Finally, additional variables can be added to the vectors, such as demographic characteristics associated with each location\cite{Kropp1998}, to undercover another dimension of associations within the resulting neighbourhood typologies.

The implications of this work are two fold, it enhances our understanding of the environments that impact our individual decisions, the systems we create and the way we are producing them as well as has concrete application in several domains impacted by the built environment.

\section{References}\label{ref}
\bibliographystyle{elsarticle-harv} 
\bibliography{BlockTypologies-Preprint}

\end{document}